\documentstyle[preprint,aps]{revtex}
\begin{document}
\draft

\title{A Hamilton-Jacobi approach to non-slow-roll inflation}
\author{William H.\ Kinney\thanks{Electronic address: {\tt kinneyw@fnal.gov}}}
\address{NASA/Fermilab Astrophysics Center}
\address{Fermi National Accelerator Labaratory, Batavia, IL 60510}
\author{FERMILAB--PUB--97/042--A}
\author{hep-ph/9702427}
\date{February 26, 1997}
\maketitle

\begin{abstract}
I describe a general approach to characterizing cosmological inflation outside the standard slow-roll approximation, based on the Hamilton-Jacobi formulation of scalar field dynamics. The basic idea is to view the equation of state of the scalar field matter as the fundamental dynamical variable, as opposed to the field value or the expansion rate.  I discuss how to formulate the equations of motion for scalar and tensor fluctuations in situations where the assumption of slow roll is not valid. I apply the general results to the simple case of inflation from an ``inverted'' polynomial potential, and to the more complicated case of hybrid inflation.
\end{abstract}


\section{Introduction}

Cosmological inflation postulates that there was a period in the very early universe in which the expansion of the universe was accelerating, so that the universe evolved toward its currently observed state of flatness and homogeneity on large scales. Inflation was first proposed by Guth in 1981\cite{guth81}, and has since been incorporated into a wide variety of models. Typical inflation models involve a single scalar field displaced from the minimum of a potential, creating a nonzero vacuum energy which dominates the stress energy of the universe. This in turn leads to quasi-exponential expansion. Inflation ends when the scalar field relaxes to the minimum of the potential and the vacuum energy decays via particle production, resulting in massive production of entropy. The equations of motion for the field are not in general analytically solvable, and approximate methods are required. By far the most widely used approximation is so-called {\it slow roll}\cite{linde82,albrecht82}, in which the evolution of the field is assumed to be strongly dominated by drag from the cosmological expansion. The slow-roll approximation not only allows analytic solution of the equation of motion of the field, but also makes possible an accurate characterization of quantum fluctuations in the field. In the inflation scenario, quantum fluctuations are responsible for the creation of small density perturbations in the early universe. These primordial perturbations act as seeds for structure formation, and are directly observable as temperature fluctuations in the cosmic microwave background (CMB) radiation. 

Slow roll is not, however, the only possibility for successfully implementing models of inflation, and solutions outside the slow-roll approximation have been found in particular situations\cite{stewart93,bellido96a}. In this paper I discuss a general approach to finding inflationary solutions outside the slow-roll approximation, based on the idea of treating the equation of state of the scalar field matter as the fundamental quantity in the dynamical equations, as opposed to the field itself. Such an approach is closely related to the {\it Hamilton-Jacobi} formalism, in which the expansion rate is treated as the dynamical variable.
A non-slow-roll solution is useful for calculating the end of inflation in models with ``inverted'' potentials, $V\left(\phi\right) = \Lambda^4 \left[1 - \left(\phi / \mu\right)^p\right]$. In these models, slow roll is a good approximation early on in inflation, $\phi \ll \mu$, but breaks down well before the end of inflation.
I also apply the formalism to the more complicated case of hybrid inflation, in which the slow-roll approximation breaks down at {\em all} points in the evolution of the scalar field.  (Non-slow-roll solutions in hybrid inflation were studied recently by Garc\'\i a-Bellido and Wands\cite{bellido96a}; I derive an equivalent result.) Finally, I discuss the general question of calculating the amplitude of metric fluctuations outside the slow-roll approximation, and derive an analytic solution for the hybrid case.

\section{The Hamilton-Jacobi Formalism}
\label{sechjreview}

In this section I review the basics of scalar field dynamics in inflationary cosmology with emphasis on the very useful {\it Hamilton-Jacobi} formalism\cite{grishchuk88,muslimov90,salopek90}. The emphasis is pedagogical; a more formal review can be found in Ref. \cite{lidsey95}. The first basic ingredient is a cosmological metric, which I shall take to be of the flat Robertson-Walker form
\begin{equation}
ds^2 = dt^2 - a^2\left(t\right) \left|d {\bf x}\right|^2 = a^2\left(\tau\right) \left[d\tau^2 - \left|d {\bf x}\right|^2\right].
\end{equation}
The quantity $\tau$ is the {\it conformal time}, with $dt = a d\tau$. The second ingredient is a spatially homogenous scalar field $\phi$ with potential $V\left(\phi\right)$ and equation of motion
\begin{equation}
\ddot \phi + 3 H \dot \phi + V'\left(\phi\right) = 0,\label{eqequationofmotion}
\end{equation}
where the {\it Hubble parameter} $H$ is defined as
\begin{equation}
H \equiv \left({\dot a \over a}\right).
\end{equation}
An overdot denotes a derivative with respect to the coordinate time $t$. If the stress energy of the universe is dominated by the scalar field $\phi$, the Einstein field equations for the evolution of the background metric $G_{\mu\nu} = 8 \pi G T_{\mu\nu}$ can be written as
\begin{equation}
H^2 = \left({\dot a \over a}\right)^2 = {2 \over 3 \kappa^2} \left[V\left(\phi\right) + {1 \over 2} \dot\phi^2\right]\label{eqbackgroundequation1}
\end{equation}
and
\begin{equation}
\left({\ddot a \over a}\right) = {2 \over 3 \kappa^2} \left[V\left(\phi\right) - \dot\phi^2\right].\label{eqbackgroundequation2}
\end{equation}
The constant $\kappa$ is defined as
\begin{equation}
\kappa^2 \equiv {m_{\rm Pl}^2 \over 4 \pi},
\end{equation}
where $m_{\rm Pl} = G^{-1/2} \simeq 10^{19}\,{\rm GeV}$ is the Planck mass. These background equations, along with the equation of motion (\ref{eqequationofmotion}), form a coupled set of differential equations describing the evolution of the universe. The fundamental quantities to be calculated are $\phi\left(t\right)$ and $a\left(t\right)$, and the potential $V\left(\phi\right)$ is input from some model. {\it Inflation} is defined to be a period of accelerated expansion 
\begin{equation}
\left(\ddot a \over a\right) > 0,\label{eqconditionforinflation}
\end{equation}
indicating an equation of state in which vacuum energy dominates over the kinetic energy of the field $\dot\phi^2 < V\left(\phi\right)$. In the limit that $\dot\phi = 0$, the expansion of the universe is of the de Sitter form, with the scale factor increasing exponentially in time:
\begin{eqnarray}
&&H = \sqrt{\left({2 \over 3 \kappa^2}\right) V\left(\phi\right)} = {\rm const},\cr
&&a \propto e^{H t}.
\end{eqnarray}
Note that with the Hubble distance $H^{-1}$ constant and the scale factor increasing exponentially, comoving length scales initially smaller than the horizon are rapidly redshifted outside the horizon. In general, the Hubble parameter $H$ will not be exactly constant, but will vary as the field $\phi$ evolves along the potential $V\left(\phi\right)$. A convenient approach to the more general case is to express the Hubble parameter directly as a function of the field $\phi$ instead of as a function of time, $H = H\left(\phi\right)$. This is consistent as long as $t$ is a single-valued function of $\phi$. Differentiating Eq. (\ref{eqbackgroundequation1}) with respect to time,
\begin{eqnarray}
2 H\left(\phi\right) H'\left(\phi\right) \dot\phi =&& \left({2 \over 3 \kappa^2}\right) \left[V'\left(\phi\right) + \ddot\phi\right] \dot \phi\cr
=&& - \left({2 \over \kappa^2}\right) H\left(\phi\right) \dot\phi^2.
\end{eqnarray}
The equation of motion (\ref{eqequationofmotion}) was used to simplify the right-hand side. Substituting back into the definition of $H$ in Eq. (\ref{eqbackgroundequation1}) results in the system of two first-order equations
\begin{eqnarray}
&&\dot\phi = -\kappa^2 H'\left(\phi\right),\cr
&&\left[H'\left(\phi\right)\right]^2 - {3 \over \kappa^2} H^2\left(\phi\right) = - {2 \over \kappa^4} V\left(\phi\right).\label{eqbasichjequations}
\end{eqnarray}
These equations are completely equivalent to the second-order equation of motion ({\ref{eqequationofmotion}). The second of these is referred to as the {\it Hamilton-Jacobi} equation, and can be written in the useful form
\begin{equation}
H^2\left(\phi\right) \left[1 - {1\over 3} \epsilon\left(\phi\right)\right] = \left({2 \over 3 \kappa^2}\right) V\left(\phi\right),\label{eqhubblehamiltonjacobi}
\end{equation}
where the parameter $\epsilon$ is defined as
\begin{equation}
\epsilon \equiv \kappa^2 \left({H'\left(\phi\right) \over H\left(\phi\right)}\right)^2.\label{eqdefofepsilon}
\end{equation}
The physical meaning of the parameter $\epsilon$ can be seen by expressing Eq.  (\ref{eqbackgroundequation2}) as
\begin{equation}
\left({\ddot a \over a}\right) = H^2 \left(\phi\right) \left[1 - \epsilon\left(\phi\right)\right],
\end{equation}
so that the condition for inflation (\ref{eqconditionforinflation}) is given simply by $\epsilon < 1$. Equivalently, $\epsilon$ can be viewed as parametrizing the equation of state of the scalar field matter, with the pressure $p$ and energy density $\rho$ related as
\begin{equation}
p = \rho \left({2 \over 3} \epsilon - 1\right).\label{eqeqnofstate}
\end{equation}
The condition for inflation $\epsilon < 1$ is the same as $\rho + 3 p < 0$.
The de Sitter case is $\epsilon = 0$ or $p = -\rho$. The evolution of the scale factor is given by the general expression
\begin{equation}
a \propto \exp\left[\int_{t_0}^{t}{H\,dt}\right],
\end{equation}
where the number of e-folds $N$ is defined to be
\begin{equation}
N \equiv \int_{t_0}^{t}{H\,dt} = \int_{\phi_0}^{\phi}{{H \over \dot\phi}\,d\phi} = \pm {1 \over \kappa} \int_{\phi_0}^{\phi}{d\phi \over \sqrt{\epsilon\left(\phi\right)}}.
\end{equation}
It is conventional to take $\phi_0$ to be the {\it end} of inflation, with $N$ increasing as one goes backward in time. In what follows, it will be convenient to define the additional parameters\cite{copeland93,liddle94}
\begin{equation}
\eta \equiv \kappa^2 \left({H''\left(\phi\right) \over H\left(\phi\right)}\right) = \epsilon \pm \left({\kappa \over 2}\right) {\epsilon' \over \sqrt{\epsilon}}\label{eqdefofeta}
\end{equation}
and
\begin{equation}
\xi \equiv \kappa^2 \left({H'\left(\phi\right) H'''\left(\phi\right) \over H^2\left(\phi\right)}\right)^{1/2} = \left[\epsilon \eta \pm \kappa \sqrt{\epsilon} \eta'\right]^{1/2}.
\end{equation}
These are often referred to as {\it slow-roll} parameters. The sign ambiguity is a result of the convention that $\sqrt{\epsilon}$ is taken to always be positive, but $H'\left(\phi\right)$ can be of either sign. The sign is fixed by specifying the sign of $\dot \phi$ in Eq. (\ref{eqbasichjequations}). The slow-roll approximation is the assumption that the field evolution is dominated by drag from the expansion $\ddot\phi \simeq 0$, so that $\dot\phi$ is approximately constant and $H\left(\phi\right)$ can be taken to vary as
\begin{equation}
H\left(\phi\right) = \sqrt{\left({2 \over 3 \kappa^2}\right) V\left[\phi\left(t\right)\right]},
\end{equation}
where $\phi\left(t\right)$ satisfies
\begin{equation}
\dot\phi = - {V'\left(\phi\right) \over 3 H\left(\phi\right)}.
\end{equation}
This approximation is consistent as long as the first two derivatives of the potential are small relative to its magnitude $V',\ V'' \ll V$. The parameters $\epsilon$ and $\eta$ reduce in this limit to\cite{kolb94} 
\begin{eqnarray}
&&\epsilon = {\kappa^2 \over 4} \left({V'\left(\phi\right) \over V\left(\phi\right)}\right)^2,\cr
&&\eta = {\kappa^2 \over 2}\left[{V''\left(\phi\right) \over V\left(\phi\right)} - {1 \over 2} \left({V'\left(\phi\right) \over V\left(\phi\right)}\right)^2\right].
\end{eqnarray}
The slow-roll limit can then be equivalently expressed as $\epsilon,\ \left|\eta\right| \ll 1$. These expressions are frequently taken in the literature as {\it definitions} of the slow-roll parameters, but here they are simply limits of the defining expressions (\ref{eqdefofepsilon}) and (\ref{eqdefofeta}). 

In the next section, I discuss the question of finding approximate solutions to the Hamilton-Jacobi equations, with particular emphasis on solutions outside the range in which slow roll is a valid approximation.

\section{Solving the Hamilton-Jacobi Equation}
\label{secgeneralformalism}

Instead of treating the Hubble parameter $H\left(\phi\right)$ as the fundamental quantity in the Hamilton-Jacobi equation, it is convenient to use  $\epsilon\left(\phi\right)$, which amounts to directly solving for the equation of state of the scalar field (\ref{eqeqnofstate}). The definition of $\epsilon$,
\begin{equation}
\epsilon\left(\phi\right) \equiv \kappa^2 \left({H'\left(\phi\right) \over H\left(\phi\right)}\right)^2,
\end{equation}
can be viewed as a differential equation which can be inverted to obtain $H\left(\epsilon\right)$, provided a boundary condition is specified. For the case of inflation, the relevant boundary condition is de Sitter expansion
\begin{equation}
\epsilon\left(\phi_0\right) \equiv 0
\end{equation}
for some field value $\phi_0$. From the definition of $\epsilon$, this is the same as saying $H'\left(\phi\right) = 0$, from which it follows that $\phi_0$ is a stationary point of the field:
\begin{equation}
\dot\phi\big|_{\phi = \phi_0} = - \kappa^2 H'\left(\phi_0\right) = 0.
\end{equation}
Subject to this boundary condition, the Hubble parameter is then {\it exactly}
\begin{equation}
H\left(\phi\right) = \sqrt{{2 \over 3 \kappa^2} V\left(\phi_0\right)} \exp\left(\pm {1 \over \kappa} \int_{\phi_0}^{\phi}{\sqrt{\epsilon\left(\phi'\right)}\,d\phi'}\right).\label{eqhofepsilon}
\end{equation}
The sign ambiguity, as in the case of the slow-roll parameters, arises as a result of the convention that $\sqrt{\epsilon}$ is taken to always be positive. The Hamilton-Jacobi equation is then, also exactly,
\begin{equation}
{V\left(\phi\right) \over V\left(\phi_0\right)} = \exp\left(\pm {2 \over \kappa}\int_{\phi_0}^{\phi}{\sqrt{\epsilon\left(\phi'\right)}\,d\phi'}\right) \left[1 - {1\over3} \epsilon\left(\phi\right)\right].\label{eqintegralhamiltonjacobi}
\end{equation}
This at first may appear to be a cumbersome form of the equation of motion, but it is in fact straightforward to recover the standard slow-roll solution by making the approximation
\begin{equation}
1 - {1 \over 3} \epsilon\left(\phi\right) \simeq 1.\label{eqslowrollapprox}
\end{equation}
The Hamilton-Jacobi equation reduces to
\begin{equation}
\int_{\phi_0}^{\phi}{\sqrt{\epsilon\left(\phi'\right)}\,d\phi'} \simeq \pm {\kappa \over 2} \ln\left({V\left(\phi\right) \over V\left(\phi_0\right)}\right),
\end{equation}
which is simply solved for $\epsilon$:
\begin{equation}
\sqrt{\epsilon\left(\phi\right)}\,d\phi = \pm {\kappa \over 2} \,d\left[\ln V\left(\phi\right)\right] = \pm {\kappa \over 2} \left({V'\left(\phi\right) \over V\left(\phi\right)}\right)\,d\phi,
\end{equation}
giving the usual slow-roll expression
\begin{equation}
\epsilon\left(\phi\right) = {\kappa^2 \over 4} \left({V'\left(\phi\right) \over V\left(\phi\right)}\right)^2.\label{eqepsilonslowroll}
\end{equation}
The Hubble parameter $H\left(\phi\right)$ is, from Eq. (\ref{eqhofepsilon}),
\begin{eqnarray}
H\left(\phi\right) =&& \sqrt{\left({2 \over 3 \kappa^2}\right) V\left(\phi_0\right)} \exp\left(\pm {1 \over \kappa} \int_{\phi_0}^{\phi}{\sqrt{\epsilon\left(\phi'\right)}\,d\phi'}\right)\cr
=&& \sqrt{\left({2 \over 3 \kappa^2}\right) V\left(\phi\right)},
\end{eqnarray}
consistent with the result obtained from directly substituting the approximation (\ref{eqslowrollapprox}) into Eq. (\ref{eqhubblehamiltonjacobi}). There is nothing new here: this is the standard slow-roll approximation, arrived at in a somewhat unconventional way. However, the form of Eq. (\ref{eqintegralhamiltonjacobi}) is suggestive of a natural extension of this approximation.  In inflation, when $\epsilon < 1$, the integral in Eq. (\ref{eqintegralhamiltonjacobi}) satisfies the inequality
\begin{equation}
\int_{\phi_0}^{\phi}{\sqrt{\epsilon\left(\phi'\right)}\,d\phi'} < \phi - \phi_0.
\end{equation}
Then if the displacement of the field from its stationary value is small compared to the Planck scale $\phi - \phi_0 \ll \kappa$, the exponential can be expanded as
\begin{equation}
\exp\left(\pm {2 \over \kappa}\int_{\phi_0}^{\phi}{\sqrt{\epsilon\left(\phi'\right)}\,d\phi'}\right) \simeq 1 \pm {2 \over \kappa}\int_{\phi_0}^{\phi}{\sqrt{\epsilon\left(\phi'\right)}\,d\phi'},
\end{equation}
and the Hamilton-Jacobi equation becomes
\begin{equation}
{V\left(\phi\right) \over V\left(\phi_0\right)} \simeq 1 - {1 \over 3} \epsilon\left(\phi\right)  \pm {2 \over \kappa}\int_{\phi_0}^{\phi}{\sqrt{\epsilon\left(\phi'\right)}\,d\phi'}.\label{eqnonslowrollintegral}
\end{equation}
In the $\kappa \rightarrow \infty$ limit, this simplifies directly to an expression for $\epsilon\left(\phi\right)$:
\begin{equation}
\epsilon\left(\phi\right) = 3 \left( 1 - {V\left(\phi\right) \over V\left(\phi_0\right)} \right).\label{eqquasidesittersolution}
\end{equation}
This is of a very different form than the familiar slow-roll expression (\ref{eqepsilonslowroll}). 

An immediate application of this solution can be found in calculating the end of inflation in models with a potential of the form
\begin{equation}
V\left(\phi\right) = \Lambda^4 \left[1 - \left({\phi \over \mu}\right)^p\right],
\end{equation}
where the ``width'' of the potential is taken to be much smaller than the Planck scale $\mu \ll \kappa$. In such models, the slow-roll approximation breaks down well before the end of inflation\cite{kinney96a,kinney96b}. Choosing the boundary condition of $\dot\phi = 0$ at the origin,  $\epsilon\left(\phi_0 \equiv 0\right) = 0$,  the solution (\ref{eqquasidesittersolution}) is an excellent approximation outside the regime in which slow roll is valid, with
\begin{equation}
\epsilon\left(\phi\right) = 3 \left(\phi \over \mu\right)^p.
\end{equation}
The end of inflation is then at $\phi_{\rm END}$ such that  $\epsilon\left(\phi_{\rm END}\right) \equiv 1$, and
\begin{equation}
\left({\phi_{\rm END} \over \mu}\right) = \left({1 \over 3}\right)^{1/p}. 
\end{equation}
This can be compared to the slow-roll result
\begin{equation}
\left({\phi_{\rm END} \over \mu}\right) = \left({2 \mu \over p \kappa}\right)^{1 / \left(p - 1\right)} \ll 1.
\end{equation}
Using the slow-roll solution for $\epsilon$ results in greatly underestimating the field value at which inflation ends. The breakdown of slow roll can be seen by calculating the parameter $\eta$ (\ref{eqdefofeta}),
\begin{equation}
\eta =  3 \left(\phi \over \mu\right)^p - {p \sqrt{3} \over 2} \left({\kappa \over \mu}\right) \left({\phi \over \mu}\right)^{p / 2 - 1},
\end{equation}
so that $\left|\eta\right|$ becomes large before the end of inflation, and slow roll is a poor approximation. However, at least for $p > 2$, the parameter $\left|\eta\right|$ becomes small and slow roll is valid for $\phi \ll \mu$, which is the region of interest when calculating observable parameters such as the spectrum of curvature fluctuations. In the next section, I discuss the more complicated case of hybrid inflation, in which slow roll breaks down in regions of interest for the calculation of observable parameters.

\section{Application to hybrid inflation}
\label{sechybridmodel}

In this section, I apply the formalism outlined in Sec. \ref{secgeneralformalism} to the case of hybrid inflation\cite{linde91,linde94,copeland94}, with a potential typically taken to be of the form
\begin{equation}
V\left(\phi,\psi\right) = \left(M^2 - {\sqrt{\lambda} \over 2} \psi^2\right)^2 + {1 \over 2} m^2 \phi^2 + {1 \over 2} g \phi^2 \psi^2.\label{eqtypicalhybridpotential}
\end{equation}
In the hybrid inflation scenario, the inflaton is the field $\phi$. The second field $\psi$ has the characteristic that its classical minimum depends on the value of the field $\phi$: for $\phi$ small, $\psi$ has a minimum at $\psi^2 = 2 M^2 / \sqrt{\lambda}$, but for $\phi$ large, the minimum is at $\psi = 0$. The transition between the two behaviors occurs at a critical value
\begin{equation}
\phi_c^2 \equiv {2 M^2 \sqrt{\lambda} \over g}.
\end{equation}
For $\phi > \phi_c$, the field $\psi$ sits at the origin, and the potential has a nonzero vacuum energy which drives inflation. The field $\phi$ is effectively the only degree of freedom, and the potential reduces to
\begin{equation}
V\left(\phi\right) = M^4 + {1 \over 2} m^2 \phi^2.\label{eqhybronefieldpotential}
\end{equation}
At $\phi = \phi_c$, the field $\psi$ becomes unstable and inflation ends. It is in principle possible for a {\it second} phase of inflation to occur as $\psi$ evolves to its minimum, with physically interesting results\cite{randall95,bellido96b}. For simplicity, I restrict myself to the case where inflation ends abruptly at $\phi = \phi_c$. In this case, it is consistent to ignore the second field altogether and simply study the formal solutions using the potential (\ref{eqhybronefieldpotential}). The appropriate inflationary boundary condition is such that $\dot\phi \rightarrow 0$ as $\phi \rightarrow 0$, or
\begin{equation}
H'\left(\phi_0 \equiv 0\right) = 0.\label{eqhybrbc}
\end{equation}
This describes a situation in which the universe inflates indefinitely, with the field asymptotically coming to rest at the origin. The true end of inflation at $\phi_c$ then serves as an arbitrary cutoff. The Hamilton-Jacobi equation (\ref{eqnonslowrollintegral}) is, with the appropriate choice of sign reflecting $\dot\phi < 0$ for $\phi > 0$,
\begin{equation}
{2 \over \kappa}\int_{0}^{\phi}{\sqrt{\epsilon\left(\phi'\right)}\,d\phi'} - {1 \over 3} \epsilon\left(\phi\right) = \left({1 \over 2}\right) {m^2 \over M^4} \phi^2,
\end{equation}
with the simple solution
\begin{equation}
\epsilon\left(\phi\right) = \left({r_{\pm} \over \kappa}\right)^2 \phi^2,
\end{equation}
where
\begin{equation}
r_{\pm} \equiv {3 \over 2} \left[1 \mp \sqrt{1 - {2 \over 3} \left({m^2 \kappa^2 \over M^4}\right)}\right].
\end{equation}
This is exactly the solution of Garc\'\i a-Bellido and Wands\cite{bellido96a}, and I am adopting their sign convention for $r_{\pm}$. A similar solution for an ``inverted'' potential, with $m^2 \rightarrow -m^2$, was obtained by Stewart and Lyth\cite{stewart93}. The parameter $\epsilon$ is by definition positive, which results in the condition
\begin{equation}
{m^2 \kappa^2 \over M^4} < {3 \over 2}.
\end{equation}
(Parameter ranges that do not satisfy this condition lead to oscillatory solutions\cite{bellido96a}.) The constant $r_{\pm}$ is easily seen to be just the $\phi \ll \kappa$ limit of the parameter $\eta$ defined in Eq. (\ref{eqdefofeta}),
\begin{eqnarray}
\eta =&& \epsilon + \left({\kappa \over 2}\right) {\epsilon' \over \sqrt{\epsilon}}\cr
=&& \left({r_{\pm} \over \kappa}\right)^2 \phi^2 + r_{\pm}\cr
\simeq&& r_{\pm},
\end{eqnarray}
so that $\epsilon$ can be expressed simply as
\begin{equation}
\epsilon\left(\phi\right) = \eta^2 \left({\phi \over \kappa}\right)^2.
\end{equation}
The slow-roll limit is $m \kappa \ll M^2$, where the parameter $\eta$ reduces to
\begin{equation}
\eta_{\rm SR} = r_{+} \simeq {1 \over 2} \left({m^2 \kappa^2 \over M^4}\right).
\end{equation}
An expression for the evolution of the field $\phi$ can be obtained by calculating the number of e-folds of inflation as the field evolves from $\phi$ to some arbitrarily chosen $\phi_1$:
\begin{equation}
N = {1 \over \kappa} \int_{\phi_1}^{\phi}{d\phi' \over \sqrt{\epsilon\left(\phi'\right)}}
= {1 \over \eta} \ln\left({\phi \over \phi_1}\right),\label{eqnumefoldshyb}
\end{equation}
so that
\begin{equation}
\phi \propto e^{\eta N}.
\end{equation}
The choice of $\phi_1$ simply amounts to the definition of where the number of e-folds $N$ vanishes, with $N > 0$ for $\phi > \phi_1$. Conventionally, this is taken to be the end of inflation, $\phi_1 = \phi_c$. The most general solution is a linear combination of $\eta = r_{\pm}$ modes, and the two solutions can be interpreted as asymptotic limits of the field evolution. The late-time limit is $\phi \rightarrow 0$, or $N \rightarrow -\infty$, and the asymptotic solution in this limit is $\phi \propto \exp\left(r_+ N\right)$. The solution $\eta = r_-$ corresponds to the $N \rightarrow +\infty$ limit. It is to be expected that the late-time limit will be the one of physical interest, since the field evolution always relaxes to this attractor after enough inflation has taken place. (One possible exception is the case in which the total amount of inflation is small, such as in ``open'' hybrid inflation\cite{bellido97}.) 

In Ref.\cite{bellido96a}, Garc\'\i a-Bellido and Wands calculate the spectrum of density fluctuations in the $\eta = r_+$ limit, assuming that the gravitational backreaction of the field can be neglected. In the next section, I calculate the spectrum of scalar density fluctuations explicitly including the gravitational backreaction, for both the limits $\eta = r_{+}$ and $\eta = r_{-}$.

\section{Density Fluctuations Far from Slow Roll}
\label{secfluctuations}

The primary observational test of inflation is observation of the cosmic microwave background (CMB) radiation. Temperature fluctuations in the CMB can be related to perturbations in the metric at the surface of last scattering. In the inflation scenario, metric perturbations are created by field fluctuations during inflation\cite{hawking82,starobinsky82,guth82,bardeen83}. During the inflationary epoch, quantum fluctuations on small scales are rapidly redshifted to scales much larger than the horizon size.  
The metric perturbations created during inflation are of two types: scalar, or {\it curvature} perturbations, which couple to the stress energy of matter in the universe and form the ``seeds'' for structure formation, and tensor, or gravitational wave perturbations, which do not couple to matter. Both scalar and tensor perturbations contribute to CMB anisotropy. Scalar fluctuations can also be interpreted as fluctuations in the density of the matter in the universe. The power spectrum of curvature perturbations is given by\cite{mukhanov92}
\begin{equation}
P_{\cal R}^{1/2}\left(k\right) = \sqrt{k^3 \over 2 \pi^2} \left|{u_k \over z}\right|
\end{equation}
where $k$ is a comoving wave number, and the mode function $u_k$ satisfies the differential equation\cite{stewart93,mukhanov85,mukhanov88}
\begin{equation}
{d^2 u_k \over d\tau^2} + \left(k^2 - {1 \over z} {d^2 z \over d\tau^2} \right) u_k = 0.\label{eqexactmodeequation}
\end{equation}
The quantity $z$ is defined as
\begin{equation}
z \equiv {a \dot\phi \over H} = - \kappa a \sqrt{\epsilon},
\end{equation}
and
\begin{equation}
{1 \over z} {d^2 z \over d\tau^2} = 2 a^2 H^2 \left(1 + \epsilon - {3 \over 2} \eta + \epsilon^2 - 2 \epsilon \eta + {1 \over 2} \eta^2 + {1 \over 2} \xi^2\right).
\end{equation}
Solutions to the second-order differential equation for the mode $u_k$ in general contain two integration constants which can be taken to be phase and normalization. The phase is fixed by the boundary condition that the mode be wavelike at short wavelengths relative to the horizon size
\begin{equation}
u_k \propto e^{- i k \tau},\ k \rightarrow \infty.\label{largekbc}
\end{equation}
The long wavelength limit $k \rightarrow 0$ is just $u_k \propto z$. Normalization is fixed by the canonical quantization condition for the fluctuations, which in terms of the $u_k$ is a Wronskian condition
\begin{equation}
u_k^{*} {d u_k \over d\tau} - u_k {d u_k^{*} \over d\tau} = - i.\label{eqnormalizationcondition}
\end{equation}

The usual method of obtaining solutions to the mode equation (\ref{eqexactmodeequation}) is to solve for the quantity $\left(a H\right)$ as a function of the conformal time $\tau$. To do this, take the exact relation
\begin{equation}
d\tau = {d\left(a H\right) \over \left(a H\right)^2 \left(1 - \epsilon\right)}
\end{equation}
and integrate by parts:
\begin{eqnarray}
\tau =&& - {1 \over \left(a H\right) \left(1 - \epsilon\right)} + \int{{d\left(a H\right) \over \left(a H\right)} {d \over d\left(a H\right)} \left(1 \over 1 - \epsilon\right)}\cr
=&& - {1 \over \left(a H\right) \left(1 - \epsilon\right)} + \int{{2 \epsilon \left(\epsilon - \eta\right) \over \left(a H\right)^2 \left(1 - \epsilon\right)^3}\,d\left(a H\right)}.\label{eqtaubyparts}
\end{eqnarray}
In the limit of power-law inflation $\epsilon = \eta = {\rm const.}$ the second integral in Eq. (\ref{eqtaubyparts}) vanishes, and the conformal time is exactly
\begin {equation}
\tau = - {1 \over \left(a H\right) \left(1 - \epsilon\right)}.
\end{equation}
The mode equation (\ref{eqexactmodeequation}) then becomes a Bessel equation, with the standard solution
\begin{equation}
u_k \propto \sqrt{- k \tau} H_\nu\left(- k \tau\right),\label{eqhankelsolution}
\end{equation}
where $H_\nu$ is a Hankel function of the first kind, and
\begin{equation}
\nu =  {3 \over 2} + {\epsilon \over 1 - \epsilon}.
\end{equation}
The limit of de Sitter expansion is $\epsilon \rightarrow 0$, and this reduces to $\nu = 3/2$, which is the case of a scale invariant spectrum $P\left(k\right) \propto k$. Thus, de Sitter expansion can be considered to be a limiting case of power-law inflation. The so-called ``slow-roll expansion'' is an expansion in small parameters about the de Sitter limit. In cases where $\epsilon \neq \eta$, but both $\epsilon$ and $\eta$ are small, the conformal time is given by the (now {\it approximate}) relation
\begin {equation}
\tau \simeq - {1 \over \left(a H\right) \left(1 - \epsilon\right)} \simeq - {1 \over \left(a H\right)} \left(1 + \epsilon\right).
\end{equation}
Note that despite the formal similarity between this and the power-law case, slow roll involves distinct assumptions, as has been pointed out recently by Grivell and Liddle\cite{grivell96}: the slow-roll and power-law solutions are the same only in the de Sitter limit. Higher-order corrections can be obtained by continuing the integration by parts,
\begin{equation}
\tau = - {1 \over \left(a H\right) \left(1 - \epsilon\right)}\left[1 + {2 \epsilon \left(\epsilon - \eta\right) \over \left(1 - \epsilon\right)^2} + {\rm O}\left(\epsilon \eta^2\right) + \cdots\right].\label{eqseriesfortau}
\end{equation}
As long as this series converges, the conformal time is well defined as a series in slow-roll parameters. In the slow-roll approximation $\epsilon,\ \eta \ll 1$, it is consistent to take  $\epsilon$ and $\eta$ to be approximately constant, and the solutions are again Hankel functions of the form (\ref{eqhankelsolution}), with 
\begin{equation}
\nu = {3 \over 2} + 2 \epsilon - \eta\label{eqslowrollnu}
\end{equation}
to first order in the slow-roll parameters. 

However, the only necessary condition for inflation is that $\epsilon$ be smaller than unity. Derivatives of $\epsilon$, in particular $\eta$, need not necessarily be small for an inflationary solution to exist. But for $\epsilon \ne 0$ and $\eta > 1$, the series (\ref{eqseriesfortau}) is not convergent and a new approach to solving the mode equations is required. Instead of expressing the mode equation (\ref{eqexactmodeequation}) as a differential equation in the conformal time $\tau$, it is convenient to switch variables to the wavelength of the fluctuation mode relative to the horizon size,
\begin{equation}
y \equiv  \left({k \over a H}\right) \simeq \left({d_H \over \lambda}\right).
\end{equation}
Then
\begin{equation}
dy = - k {d \left(a H\right) \over \left(a H\right)^2}
= - k \left(1 - \epsilon\right) d\tau,\label{eqdifferentialy}
\end{equation}
and the mode equation (\ref{eqexactmodeequation}) can be expressed exactly as
\begin{equation}
y^2 \left(1 - \epsilon\right)^2 {d^2 u_k \over d y^2} + 2 y \epsilon \left(\epsilon - \eta\right) {d u_k \over d y} + \left[y^2 - F\left(\epsilon, \eta, \xi\right)\right] u_k = 0,\label{eqexactmodeequationy}
\end{equation}
where 
\begin{equation}
F\left(\epsilon,\eta,\xi\right) \equiv 2 \left(1 + \epsilon - {3 \over 2} \eta + \epsilon^2 - 2 \epsilon \eta + {1 \over 2} \eta^2 + {1 \over 2} \xi^2\right).
\end{equation}
The approximately deSitter limit $\epsilon \ll 1$, $\eta = {\rm const} \gg \epsilon$ is just the case of hybrid inflation considered in Sec. \ref{sechybridmodel}. In this case, the mode equation (\ref{eqexactmodeequationy}) reduces to
\begin{equation}
y^2 {d^2 u_k \over d y^2} + \left[y^2 - \left(2 + \eta^2 - 3 \eta\right)\right] u_k = 0.
\end{equation}
The solution is again a Bessel function:
\begin{eqnarray}
&&u_k \propto y^{1/2} H_\nu\left(y\right),\cr
&&\nu = {3 \over 2} - \eta,
\end{eqnarray}
in agreement with the calculation in Ref. \cite{bellido96a}. This expression is the same as the $\epsilon \rightarrow 0$ limit in the slow-roll case (\ref{eqslowrollnu}), although here no assumption of small $\eta$ has been made. This is to be expected, since for $\epsilon$ exactly vanishing, the series (\ref{eqseriesfortau}) converges for any $\eta$. For $\epsilon$ nonzero but small, the series is asymptotic, and it is still a consistent approximation to take $y \simeq -k \tau$ as suggested by the differential relationship (\ref{eqdifferentialy}).  Note in particular that $\nu$ changes sign when $\eta = 3/2$, which is equivalent to a change of {\it phase} in the solution. The phase of the fluctuation is set by the boundary condition (\ref{largekbc}), so the solutions $\nu = \pm 3/2$ represent physically the same fluctuation mode. For the case of hybrid inflation, the significance of the two solutions $\eta = r_{\pm}$ is now clear, since
\begin{equation}
\nu = {3 \over 2} - \eta = \pm {3 \over 2} \sqrt{1 - {2 \over 3} \left({m^2 \kappa^2 \over M^4}\right)}.
\end{equation}
The solutions $\eta = r_{+}$ and $\eta = r_{-}$ produce identical fluctuation modes, and the order of the Hankel function can be taken to be $\nu = \left|3/2 - \eta\right|$ without loss of generality. The mode function normalized according to the condition (\ref{eqnormalizationcondition}) is
\begin{equation}
\left|u_k\right| = {1 \over 2} \sqrt{\pi \over k} \left|y^{1/2} H_\nu\left(y\right)\right|.
\end{equation}
In the long wavelength limit $y \rightarrow 0$ this reduces to
\begin{equation}
\left|u_k\right| \rightarrow 2^{\nu - 3/2} {\Gamma\left(\nu\right) \over \Gamma\left(3/2\right)} {y^{-\nu + 1/2} \over \sqrt{2 k}},\ \ \ \nu \equiv \left|{3 \over 2} - \eta\right|.
\end{equation}
The normalized fluctuation amplitude is conventionally evaluated at horizon crossing $y = \left(k / a H\right) = 1$, and is given by
\begin{eqnarray}
P_{\cal R}^{1/2}\left(k\right) &&= {2^{\nu - 3/2} \over 2 \pi} {\Gamma\left(\nu\right) \over \Gamma\left(3/2\right)} \left({H \over \kappa \sqrt{\epsilon}}\right)\Biggr|_{k = a H}\cr
&&=  {2^{\nu - 3/2} \over 2 \pi} {\Gamma\left(\nu\right) \over \Gamma\left(3/2\right)} {e^{-\eta N} \over \eta} \left({H \over \kappa \phi_c}\right),
\end{eqnarray}
where $N$ is the number of e-folds before the end of inflation, evaluated at horizon crossing. This is typically taken to be $N \simeq 60$. The scalar spectral index is
\begin{equation}
n - 1 = {d \ln{P_{\cal R}} \over d \ln{k}} =  2 \eta.
\end{equation}
This answer reduces to the result in Ref. \cite{bellido96a} in the limit $\eta = r_+ \leq 3/2$, and confirms the consistency of neglecting the gravitational backreaction when calculating the power spectrum of curvature perturbations. In the $\eta = r_-$ limit, the spectral index becomes large, $n \geq 4$, which lies well outside the limits set by the Cosmic Background Explorer (COBE) satellite, $n = 1.2 \pm 0.3$\cite{bennett96,gorski96}. Note that although the mode function $u_k$ is the same for both limits $\eta = r_\pm$, the {\em background} evolution is different in the two limits, which leads to the difference in the spectral index.

The situation is similar for the case of tensor fluctuations. The tensor mode equation can be written
\begin{equation}
{d^2 v_k \over d\tau^2} + \left[k^2 - a^2 H^2 \left(2 - \epsilon\right)\right] v_k = 0,
\end{equation}
where the amplitude of the tensor metric perturbation is given by $v_k / a$. In terms of the variable $y$, this becomes
\begin{equation}
y^2 \left(1 - \epsilon\right)^2 {d^2 v_k \over d y^2} + 2 y \epsilon \left(\epsilon - \eta\right) {d v_k \over d y} + \left[y^2 - \left(2 - \epsilon\right)\right] v_k = 0,
\end{equation}
which can be solved in a fashion similar to the scalar mode equation (\ref{eqexactmodeequationy}). However, in the limit $\epsilon \rightarrow 0$, tensor fluctuations become negligible relative to scalar fluctuations, and I do not consider them further here.

\section{Conclusions}
\label{secconclusions}

In this paper, I have outlined a general way of approaching the problem of inflation beyond the slow-roll approximation. In this approach, the fundamental quantity is the parameter $\epsilon\left(\phi\right)$, which characterizes the equation of state of the scalar field matter $p = \rho \left(2 \epsilon / 3 - 1\right)$. The Hamilton-Jacobi equation for the evolution of a scalar field in inflation can be written in the exact form
\begin{equation}
{V\left(\phi\right) \over V\left(\phi_0\right)} = \exp\left(\pm {2 \over \kappa}\int_{\phi_0}^{\phi}{\sqrt{\epsilon\left(\phi'\right)}\,d\phi'}\right) \left[1 - {1\over3} \epsilon\left(\phi\right)\right],
\end{equation}
where $\phi_0$ is a boundary value such that $\epsilon\left(\phi_0\right) \equiv 0$. In this formulation, the standard slow-roll approximation corresponds to taking $1 - \epsilon/3 \simeq 1$, where the solution is the usual expression
\begin{equation}
\epsilon_{\rm SR} = {\kappa^2 \over 4}\left({V'\left(\phi\right) \over V\left(\phi\right)}\right)^2.
\end{equation}
This is often taken as the definition of the parameter $\epsilon$. However, in the limit $\phi - \phi_0 \ll \kappa$, a general non-slow-roll solution exists:
\begin{equation}
\epsilon = 3 \left(1 - {V\left(\phi\right) \over V\left(\phi_0\right)}\right).
\end{equation}
This expression is useful, for instance, for accurate calculation of the end of inflation in models with ``inverted'' potentials, $V = \Lambda^4\left[1 - \left(\phi / \mu\right)^p\right]$, where the slow-roll approximation gives poor results.

I apply the general formalism to the interesting example of hybrid inflation, with a potential of the form
\begin{equation}
V\left(\phi\right) = M^4 + {1\over 2} m^2 \phi^2.
\end{equation}
I derive a solution to the background field equations equivalent to that obtained by Garc\'\i a-Bellido and Wands\cite{bellido96a}:
\begin{eqnarray}
&&\epsilon = \eta^2 \phi^2,\cr
&&\phi \propto e^{\eta N},
\end{eqnarray}
where $N$ is the number of e-folds of inflation, and the parameter $\eta$ is given by
\begin{equation}
\eta \equiv \kappa^2 \left({H'\left(\phi\right) \over H\left(\phi\right)}\right)^2 = {3 \over 2}\left[1 \mp \sqrt{1 - {2 \over 3}\left({m^2 \kappa^2 \over M^4}\right)}\right] \equiv r_{\pm}.
\end{equation}
The $\eta = r_+$ solution corresponds to the late-time attractor, and the $\eta = r_-$ solution corresponds to the $N \rightarrow +\infty$ limit. The amplitude of curvature fluctuations is
\begin{equation}
P_{\cal R}^{1/2}\left(k\right) =  {2^{\nu - 3/2} \over 2 \pi} {\Gamma\left(\nu\right) \over \Gamma\left(3/2\right)} {e^{-\eta N} \over \eta} \left({H \over \kappa \phi_c}\right),\quad \nu \equiv \left|{3 \over 2} - \eta\right|,
\end{equation}
where $\phi_c$ is the field value at the end of inflation, and $N$ is the number of e-folds to the end of inflation when the fluctuation crosses the horizon. The spectral index of curvature fluctuations is
\begin{equation}
n = 1 + 2 \eta,
\end{equation}
so the $\eta = r_-$ asymptote gives an unacceptably large spectral index, $n \geq 4$. Tensor modes are negligible in all cases.

\section*{Acknowledgments}

I would like to thank Edward Kolb and Antonio Riotto for helpful conversations relating to this work. David Wands provided crucial insights regarding solutions in the hybrid inflation case.

This work was supported in part by DOE and NASA Grant No. NAG5-2788 at Fermilab.

\end{document}